\documentclass[prl,amsmath,amsfonts,amssymb,twocolumn,superscriptaddress,showpacs]{revtex4}

\usepackage{graphicx,psfrag}% Include figure files
\usepackage{dcolumn}% Align table columns on decimal point
\usepackage{bm}% bold math
\usepackage{mathbbol}

\def\openone{\leavevmode\hbox{\small1\kern-3.8pt\normalsize1}}
\def\RR{{\rm I\kern-.2emR}}

\def\openone{\leavevmode\hbox{\small1\kern-3.8pt\normalsize1}}
\def\RR{{\rm I\kern-.2emR}}

\providecommand{\ignore}[1]{}

\newcommand{\ket}[1]{| #1 \rangle}
\newcommand{\bra}[1]{\langle #1 |}

\newcommand{\bitem}{\begin{itemize}}
\newcommand{\eitem}{\end{itemize}}
\newcommand{\benum}{\begin{enumerate}}
\newcommand{\eenum}{\end{enumerate}}
\newcommand{\beq}{\begin{equation}}
\newcommand{\eeq}{\end{equation}}
\newcommand{\beqa}{\begin{eqnarray}}
\newcommand{\eeqa}{\end{eqnarray}}
\newtheorem{definition}{Definition}

\newtheorem{proposition}{Proposition}

\newcommand{\bproof}{\begin{proof}}
\newcommand{\eproof}{\end{proof}}
\newcommand{\bprop}{\begin{proposition}}

\newcommand{\bdef}{\begin{definition}}

\begin{document}
\setcounter{secnumdepth}{1}

\title{Quantum Noise Filtering via Cross-Correlations}

\author{Boaz Tamir}
\email{canjlm@actcom.co.il}
\affiliation{\mbox{Faculty of Interdisciplinary Studies, S.T.S Program, Bar-Ilan University, Ramat-Gan, Israel}\\{IYAR, Israel institute for advanced research}}

\author{Eliahu Cohen}
\email{eliahuco@post.tau.ac.il}
\affiliation{\mbox{School of Physics and Astronomy, Tel Aviv University, Tel Aviv, Israel}}

%\author{Sergei Masis}
%\email{srjmas@gmail.com}
%\affiliation{\mbox{Department of physics, Technion, Haifa, Israel}}

\date{\today}

\pacs{03.67.Ac;42.50.Lc}
\begin{abstract}
\textbf{Abstract}\\
Motivated by successful classical models for noise reduction, we suggest a quantum technique for filtering noise out of quantum states. The purpose of this paper is twofold: presenting a simple construction of quantum cross-correlations between two wave-functions, and presenting a scheme for a quantum noise filtering. We follow a well-known scheme in classical communication theory that attenuates random noise, and show that one can build a quantum analog by using non-trace-preserving operators. By this we introduce a classically motivated signal processing scheme to quantum information theory, which can help reducing quantum noise, and particularly, phase flip noise.
%One of the most common quantum noise processes is the phase damping channel.

\end{abstract}

\maketitle

\vspace{1cm}

%{\bf Key words:} ????.

\noindent

%%%%%%%%%%%%%%%%%%%%%%%%%%%%%%%%%%%%%%%%%%%%%%

\section {Introduction and motivation}
\label{sec:intro}

In classical communication theory, the use of cross-correlation and autocorrelation functions is very common. There is a family of classical algorithms for the retrieval of information below noise. These are based on cross-correlations and use some type of referential wave function to detect the presence of a signal, or its shape \cite{Champeney}, \cite{Schwartz}. To name just a few, there is the phase sensitive detector that uses a synchronous referential wave form; the Boxcar detector that correlates a repetitive waveform with a pulse function as a gating function; the matched filter that detects the presence of a signal with known shape but unknown amplitude, and the more general correlator; the lock-in amplifier; the integrator which is a low-pass filter, etc. Cross-correlations are also used in spectrum analysis or for estimating the level of randomness, etc. see \cite{Sharma} - \cite{Cover}.
Here we wish to present a quantum analog for the classical correlator, that is, a quantum noise filter that utilizes cross-correlations to attenuate the noise.

The power of cross-correlation stems from a deep connection between the correlation and the energy or power spectrum density \cite{Wiener}. In classical theory this is known as the Wiener-Khinchin theorem \cite{Champeney}. Similar relations are also true in quantum optics \cite{Garrison}.\\

Quantum correlations are well-known in quantum optics following the work of Glauber \cite{Glauber}.  The single photon interference in a two-slit experiment can be described using the first-order cross-correlation Glauber function $G^{(1)} (r_1,t_1,r_2,t_2)$. Moreover, the intensity-intensity correlation or the Hanbury-Brown-Twiss effect \cite{Hanbury} are explained via the second-order Glauber function  $G^{(2)} (r_1,t_1,r_2,t_2)$. Cross-correlations are also used in photon detection, an example could be the heterodyne or monodyne detection schemes \cite{Garrison}. Both are the quantum (optical) analogues of the classical schemes for the detection of weak radio frequency signals. For other uses of cross-correlations see for example \cite{Gerry}, \cite{Barnett}.

%We use the context of quantum information and quantum computation, and therefore the language of von-Neumann measurement theory, post-selection and operator sum representation (the Kraus representation) \cite{Nielsen}.\\

%In what follows we consider a quantum signal recorded in a quantum memory array. By applying quantum cross-correlation processing to the signal, we increase the signal to noise ratio and peel off the accumulated phase noise.  Other cross-correlation methods from classical signal processing might also be applied. The construction of quantum cross-correlation apparatus is the main innovation of this paper.

The paper is organized as follows. We start (section II) with a few essential preliminaries which will be used later on for the construction of the proposed method. The heart of the paper lies in section III. First we describe a simple way to construct correlation integrals. Given a discrete pure wave function with density matrix $\mathcal{S}$ and a `reference' (see below) discrete pure wave function $\mathcal{S}_0$ we present a non trace-preserving operator $\mathcal{E}_{\mathcal{S}_0}$ such that $\mathcal{E}_{\mathcal{S}_0}(\mathcal{S})$ is a density matrix with the correlations $\langle S,S_0 \rangle$ of $S_0$ and $S$ as its coefficients. We use von-Neuman measurement theory \cite{Neumann} and some techniques used in weak measurement theory \cite{Aharonov,ACE} to construct $\mathcal{E}$ (also discussed in the appendix).\\

Next we analyze the case of a quantum signal with noise; let $\rho= p\mathcal{S}+ (1-p)\mathcal{N}$ where $\mathcal{N}= \sum_i E_iSE_i^\dag$ and $\sum E^\dag E =1$. We prove that our operator $\mathcal{E}$ increases the fidelity between $\rho$ and $\mathcal{S}$, to be more precise:

\[ F(\mathcal{E}(\rho),\mathcal{E}(S))\geq F(\rho,\mathcal{S})\]

\noindent We also show that the increase in fidelity is paid out by the number of postselections (final projective measurements) in the construction of $\mathcal{E}_{\mathcal{S}_0}(\mathcal{S})$. We therefore reduce the amount of noise and pay with the number of measurements. The operator $\mathcal{E}_{\mathcal{S}_0}$ is the quantum parallel of a classical correlator or low-pass filter integrator. Another way to look at it, is as a lock-in amplifier (recently discussed in \cite{Kotler}). \\

In section IV we discuss the results and suggest a few future research directions.

\section {Preliminaries}

In this section we outline the scope of our problem and define the correlator which will be later used for noise filtering.

Let $\ket{\phi} = \sum_{k=1}^N \phi(k) \ket{k}$ and $\ket{\psi} = \sum_{k=1}^N \psi(k) \ket{k}$ be two pure state vectors. We will also use the notation:
$\ket{\phi(k)} =\phi(k) \ket{k}$\\

\textbf{Definition:} Define the correlation coefficient as:
\begin{eqnarray}
 C(\ket{\phi}, \ket{\psi}) = \frac{1}{N} \sum_{i=1}^N | \bra{\phi(k-i)} \psi(k) \rangle_k |^2,
\end{eqnarray}
\noindent where $\bra{\phi(k-i)} \psi(k) \rangle_k$ is the $k$ integral making the scalar product. \\

The following lemma is a simple consequence of the above definition:\\

\textbf{Lemma:} $0 \leq C(\ket{\phi}, \ket{\psi})\leq 1$.\\

Consider the following density matrix describing a signal with some noise:
\begin{eqnarray}
\rho = p \ket{\phi} \bra{\phi} + (1-p) E \ket{\phi} \bra{\phi} E^\dag,
\end{eqnarray}
\noindent where $E^\dag E = 1$. Let $\mathcal{S}$ denote the signal $\ket{\phi} \bra{\phi}$ and $\mathcal{N}$ the noise $E \ket{\phi} \bra{\phi} E^\dag$, then we can extend the above definition of correlation coefficient to the densities:
\begin{eqnarray}
C(S,N) = \frac{1}{N} \sum_{i=1}^N | \bra{\phi(k-i)} E \ket{\phi(k)}_k |^2.
\end{eqnarray}
\textbf{Example:} For a phase flip type of noise $E = \frac{1}{n}\sum_{i=1}^n Z^{(i)}$ and a fixed amplitude signal $\ket{\phi} = \frac{1}{\sqrt{N}} \sum_{k=1}^N \ket{k}$ it is easy to see that: \[C(\mathcal{S},\mathcal{N})=0\]

\[C(\mathcal{S},\mathcal{S})=1.\]

Therefore, if we could somehow correlate $\rho$ with $S$ we will eventually get rid of the noise. In other words, we are looking for a quantum counterpart of the classical scheme for filtering noise by correlation integrals. \\

Classically, if $\mathcal{S}$ is a signal, $\mathcal{N}$ is some random amplitude noise and $\langle\mathcal{S}+\mathcal{N},\mathcal{S}+\mathcal{N}\rangle$ is their autocorrelation, then:

\[ \langle\mathcal{S}+\mathcal{N},\mathcal{S}+\mathcal{N}\rangle = \langle\mathcal{S},\mathcal{S}\rangle + \langle\mathcal{S},\mathcal{N}\rangle+\]
\[+ \langle\mathcal{N},\mathcal{S}\rangle + \langle\mathcal{N},\mathcal{N}\rangle = \langle\mathcal{S},\mathcal{S}\rangle .\]

\noindent It is well-known that the spectrum of $\langle\mathcal{S},\mathcal{S}\rangle$ is very close to that of $\mathcal{S}$, and therefore we can analyze $\mathcal{S}$ by analyzing $\langle\mathcal{S},\mathcal{S}\rangle$, regardless of the noise $\mathcal{N}$ \cite{Hancock}. This scheme is not practical as a quantum process since it demands the cloning of $\mathcal{S}+\mathcal{N}$ \cite{Wootters}. Therefore we will use a similar scheme:

\[ \langle\mathcal{S}_0,\mathcal{S}+\mathcal{N}\rangle = \langle\mathcal{S}_0,\mathcal{S}\rangle + \langle\mathcal{S}_0,\mathcal{N}\rangle,\]

\noindent where $\mathcal{S}_0$ is some known referential function. \\

To construct the correlation functions we will present a non-trace-preserving operator $\mathcal{E}= \mathcal{E}_{\ket{\phi_0}}$ where $\ket{\phi_0}\bra{\phi_0}$ is some referential signal $S_0$. Applying $\mathcal{E}$ to $\rho$ we have:
\begin{eqnarray}
\mathcal{E}_{\ket{\phi_0}}(\rho) = q \mathcal{E}_{\ket{\phi_0}}(\mathcal{S})+ (1-q) \mathcal{E}_{\ket{\phi_0}}(\mathcal{N}),
\end{eqnarray}
\noindent where $\mathcal{E}_{\ket{\phi_0}}(\mathcal{S})$ (res. $\mathcal{E}_{\ket{\phi_0}}(\mathcal{N}$)) is a density matrix with the correlations $\langle{\phi_0(k-i)}, \phi(k)\rangle$ (res. $\bra{\phi_0(k-i)} E \ket{\phi(k)} $) as its coefficients. Hence the correspondence of quantum to classical signals is as follows:
\begin{eqnarray}
\end{eqnarray}
\[ \mathcal{E}_{\ket{\phi_0}}(\mathcal{S}) \sim \langle\mathcal{S}_0,\mathcal{S}\rangle\]

\[ \mathcal{E}_{\ket{\phi_0}}(\mathcal{N}) \sim \langle\mathcal{S}_0,\mathcal{N}\rangle\]

\[ \mathcal{E}_{\ket{\phi_0}}(\rho) \sim \langle\mathcal{S}_0,\mathcal{S+N}\rangle. \] \

\noindent The probability $q$ (res. $1-q$) is a functions of $p$ (res. 1-p)) and the correlation coefficient $C(\mathcal{S}_0,\mathcal{S})$ (res. $C(\mathcal{S}_0,\mathcal{N})$). In terms of fidelity measure we will further show that:
\begin{eqnarray}
F( \mathcal{E}_{\ket{\phi_0}}(\rho), \mathcal{E}_{\ket{\phi_0}}(\mathcal{S})) \geq F(\rho, \mathcal{S}),
\end{eqnarray}
\noindent and therefore $\mathcal{E}_{\ket{\phi_0}}(\rho)$ can be looked at as a rotation in the direction of the signal and away from the noise.\\

\section {The construction of correlations and autocorrelations}

In this section we will provide a general argument showing how to construct correlations and autocorrelations between wave-functions. We will follow the general scheme presented below:\\

\setlength{\unitlength}{0.75mm}
\begin{picture}(150,80)(-20,-20)
\put(0,0){\framebox(30,30)}
\put(2,23){\makebox(0,0)[bl]{$A$}}
\put(2,3){\makebox(0,0)[bl]{$Q$}}
\put(-25,27){\makebox(0,0)[bl]{$\ket{\psi}\bra{\psi}$}}
\put(-25,7){\makebox(0,0)[bl]{$\ket{\phi}\bra{\phi}$}}
\put(15,37){\makebox(0,0)[bl]{$\frac{1}{N} \sum_{i,k,j,l} \ket{\phi(k-i)}\ket{i} \bra{j} \bra{\phi(l-j)}$}}
\put(13,13){\makebox(0,0)[bl]{$U\rho U^\dag$}}
\put(40,7){\makebox(0,0)[bl]{$\hat{M}=\ket{\phi_0}\bra{\phi_0}$}}
\put(-20,5){\line(1,0){20}}
\put(-20,25){\line(1,0){20}}
\put(30,25){\line(1,0){40}}
\put(30,5){\line(1,0){5}}
\put(30,5){\line(1,0){40}}
\put(45,5){\line(1,0){10}}
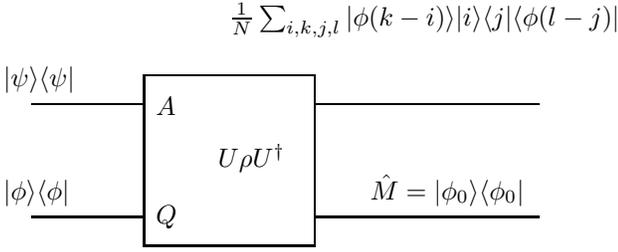
\put(-25,-15){\makebox(0,0)[bl]{{\bf Fig.1.}  Schematic illustration of the correlator  }}
\end{picture}\\

Given a system $Q$ described by the density matrix $\rho=\ket{\phi}\bra{\phi}$ of the pure state $\ket{\phi}$, and a system $A$ described by  $\ket{\psi}\bra{\psi} = \frac{1}{N} \sum_{i,j} \ket{i}\bra{j}$, we will couple the two systems by a unitary operator that will entangle them (see the Appendix for further details). Thus we will get:
\begin{eqnarray}
U\rho U^\dag = \frac{1}{N}\sum_{i,k,j,l} \ket{\phi(k-i)}\ket{i} \bra{j} \bra{\phi(l-j)}.
\end{eqnarray}

\noindent Next we will post-select the referential function $\ket{\phi_0}$ using the measurement operator $\hat{M}= \ket{\phi_0}\bra{\phi_0}$. This will leave the system $A$ as a density matrix with the cross-correlations $\bra{\phi_0(k)} \phi(k-i)\rangle_k$  as matrix coefficients. Explicitly:
\begin{eqnarray}
\end{eqnarray}
\[ \mathcal{E}\ket{\phi}\bra{\phi} = \frac{1}{N} \sum_{i,j} \bra{\phi_0(k)} \phi(k-i)\rangle_k \bra{\phi(k-j)} \phi_0(k)\rangle_k \ket{i} \bra{j}.\]

\noindent Note the following trace formula:
\begin{eqnarray}
\label{trace}
\end{eqnarray}
\[ tr(\mathcal{E}\ket{\phi}\bra{\phi}) = \frac{1}{N} \sum_i | \bra{\phi_0(k)} \phi(k-i)\rangle_k|^2 = C(\ket{\phi_0},\ket{\phi}).\]

\noindent In other words, the probability to get the post-selection $\ket{\phi_0}\bra{\phi_0}$ is the average correlation between the reference state $\ket{\phi_0}$ and the original signal $\ket{\phi}$. If the correlation between the reference and the signal is high, then the probability to post-select the reference is also high.\\

We will now show that the operator $\mathcal{E}= \mathcal{E}_{\ket{\phi_0}}$ increases the fidelity $F$.\\

\noindent \textbf{Theorem:} $F(\mathcal{E}(\rho),\mathcal{E}(S))\geq F(\rho,S)$.\\

Since $\mathcal{E}(\rho)$ is non-trace-preserving we have to normalize it by its trace. If we use the reference $\ket{\phi_0}$ then $\mathcal{E}= \mathcal{E}_{\ket{\phi_0}}$ and:
\begin{eqnarray}
\end{eqnarray}
\[ \frac{\mathcal{E}_{\ket{\phi_0}} (\rho)}{tr\mathcal{E}_{\ket{\phi_0}} (\rho)} = p \frac{\mathcal{E}_{\ket{\phi_0}} (\ket{\phi}\bra{\phi})}{tr\mathcal{E}_{\ket{\phi_0}} (\rho)}+(1-p) \frac{\mathcal{E}_{\ket{\phi_0}} (E\ket{\phi}\bra{\phi}E^\dag)}{tr\mathcal{E}_{\ket{\phi_0}} (\rho)}.\]

\noindent Next we normalize $\mathcal{E}_{\ket{\phi_0}} (\ket{\phi}\bra{\phi})$ and $\mathcal{E}_{\ket{\phi_0}} (E\ket{\phi}\bra{\phi}E^\dag)$ to get:
\begin{eqnarray}
\end{eqnarray}
\[ \frac{\mathcal{E}_{\ket{\phi_0}} (\rho)}{tr\mathcal{E}_{\ket{\phi_0}} (\rho)}= p \frac{tr \mathcal{E}_{\ket{\phi_0}}(\ket{\phi}\bra{\phi})}{tr\mathcal{E}_{\ket{\phi_0}} (\rho)}  \frac{\mathcal{E}_{\ket{\phi_0}} (\ket{\phi}\bra{\phi})}{tr\mathcal{E}_{\ket{\phi_0}} (\ket{\phi}\bra{\phi})}+\]
\[+(1-p) \frac{tr \mathcal{E}_{\ket{\phi_0}}(E\ket{\phi}\bra{\phi}E^\dag)}{tr\mathcal{E}_{\ket{\phi_0}} (\rho)}  \frac{\mathcal{E}_{\ket{\phi_0}} (E\ket{\phi}\bra{\phi}E^\dag)}{tr\mathcal{E}_{\ket{\phi_0}} (E\ket{\phi}\bra{\phi}E^\dag)}.\]

\noindent However by the above trace formula ($\ref{trace}$):
\begin{eqnarray}
\end{eqnarray}
\[ \tilde{\mathcal{E}}_{\ket{\phi_0}} (\rho)= p \frac{ C(\ket{\phi_0},\ket{\phi})}{tr\mathcal{E}_{\ket{\phi_0}} (\rho)} \cdot \tilde{\mathcal{E}}_{\ket{\phi_0}} (\ket{\phi}\bra{\phi})\]
\[+ (1-p) \frac{ C(\ket{\phi_0},E\ket{\phi})}{tr\mathcal{E}_{\ket{\phi_0}} (\rho)} \cdot
\tilde{\mathcal{E}}_{\ket{\phi_0}} (E\ket{\phi}\bra{\phi}E^\dag),\]

\noindent where $\tilde{\mathcal{E}}(\rho)$ denotes the normalization of $\mathcal{E}(\rho)$.\\

%Write now:

%\[ q= p \frac{ C(\ket{\phi_0},\ket{\phi})}{tr\mathcal{E}_{\ket{\phi_0}} (\rho)} \]

%\[ 1-q = (1-p) \frac{ C(\ket{\phi_0},\ket{E\phi})}{tr\mathcal{E}_{\ket{\phi_0}} (\rho)}\]

\noindent The density matrix $\tilde{\mathcal{E}}_{\ket{\phi_0}} (\ket{\phi}\bra{\phi})$ is now multiplied by the correlation coefficient $C(\ket{\phi_0},\ket{\phi})$, while the density matrix $\tilde{\mathcal{E}}_{\ket{\phi_0}} (E\ket{\phi}\bra{\phi}E^\dag)$ is multiplied by the correlation coefficient $C(\ket{\phi_0},E\ket{\phi})$. By the strong concavity of the fidelity (see \cite{Nielsen} theorem 9.2) we can write:
\begin{eqnarray}
 F(\mathcal{E}(\rho),\mathcal{E}(S))\geq \sqrt{ p \frac{ C(\ket{\phi_0},\ket{\phi})}{tr\mathcal{E}_{\ket{\phi_0}} (\rho)}}.
\end{eqnarray}
\noindent We can also compute $F(\rho,S)$ directly:
\begin{eqnarray}
F(\rho,S) = \sqrt{p +(1-p) \bra{\phi} E \ket{\phi}^2}.
\end{eqnarray}
\noindent Whenever $C(\ket{\phi_0},\ket{\phi})$ is close to $1$ (by choosing the right referential function) and $C(\ket{\phi_0},E \ket{\phi})$ is close to $0$ (this will depend on the type of noise) we can guarantee that:
\begin{eqnarray}
F(\rho,S)\approx \sqrt{p},
\end{eqnarray}
\noindent and
\begin{eqnarray}
F(\mathcal{E}(\rho),\mathcal{E}(S))\approx  \sqrt{ \frac{ p}{tr\mathcal{E}_{\ket{\phi_0}} (\rho)}}.
\end{eqnarray}
\hspace{75mm} $\blacksquare$\\

%It is left to show that:

%\[ \sqrt{ p \frac{ C(\ket{\phi_0},\ket{\phi})}{tr\mathcal{E}_{\ket{\phi_0}} (\rho)}}\geq \sqrt{p} \]
\textbf{Corollary:} The increase in fidelity due to the filter is proportional to:
\begin{eqnarray}
\frac{1}{\sqrt{tr\mathcal{E}_{\ket{\phi_0}} (\rho)}}.
\end{eqnarray}
\hspace{75mm} $\blacksquare$\\

It is important to note that filtering the noise has a cost in terms of of the number of post-selection trials. This is the content of the above corollary. We will need several applications of the protocol until post-selection is achieved. This argument is similar to the one employed in signal amplification protocols using weak measurement methods \cite{Amp1,Amp2}.\\

%which is formed by the convolution of $\ket{Q'(x)}$ and $\ket{Q(x)}$.\\

\textbf{Example:} For $\ket{\phi}= \ket{\phi_0}=\frac{1}{\sqrt{N}} \sum_i \ket{i}$ and $E=\frac{1}{n}\sum_i Z^{(i)}$ a phase flip noise as above:

\[ F(\rho,S)= \sqrt{p}\]

\[ F(\mathcal{E}(\rho),\mathcal{E}(S))=1,\]

\noindent and the increase in fidelity is exactly $\frac{1}{\sqrt{p}}$.

\section{Discussion}

Motivated by the theory of cross-correlations in classical physics and its successful applications such as filters, correlators, integrators, etc. we extended its methods to the theory of quantum information processing. \\

We have constructed a simple method for creating cross-correlations using a quantum measurement scheme followed by post-selection. The cross-correlation integrals (for each of the lags) are represented as the amplitudes of the output vector of the quantum filter, whereas the average cross correlation over all lags corresponds to the post-selection probability. This immediately suggests the construction of a quantum noise filter. We have shown that such a filter can be defined by the use of a non-trace-preserving operator.\\

For the protocol to work we need to be sure that the correlation of the reference function with the original signal is high, and the type of noise is such that the correlation of the noise with the signal is low. These conditions are similar to those in the classical counterpart of lock-in amplifier.

We applied the proposed method to the case of phase flip channel in the special case there the input signal was constant. The phase flip channel is strictly related to the phase damping channel- one of the most subtle and important processes in the study of quantum information. The use of cross-correlation led to a significant reduction of this noise. \\

We believe that the above scheme (or a modification of it) could be generalized to different wave-functions and noise patterns. Moreover, this suggests the use of other classical signal processing techniques in quantum information theory.

\section{Appendix}

We shall describe the entanglement process made by $U$ in details, first for a continuous, and then for an almost discrete, quantum pointers.
Consider a system $Q$ described by a the density matrix $\ket{\eta}\bra{\eta}$. Let $V_Q$ be its Hilbert space and let $\ket{\eta}$ be described by a real continuous wave-function:
\begin{eqnarray}
\eta(x)= \left(\frac{1}{2\pi\sigma^2}\right)^{1/4} e^{-\frac{x^2}{4\sigma^2}}
\end{eqnarray}

%\noindent where $\hat{X} \ket{x}= x \ket{x}$. Let $\hat{P}$ be the operator conjugate to $\hat{X}$, such that $[\hat{X},\hat{P}]= i\hbar$. \\

We couple $Q$ to another system $A$ with $N$ eigenvectors $\ket{a_j}$ such that $\hat{A}\ket{a_j}= a_j \ket{a_j}$. The eigenvectors $\ket{a_j}$ will be used to shift the argument of the function $\eta(x)$. Consider also the state vector

\[ \ket{\psi} = \frac{1}{\sqrt{N}} \sum_j \ket{a_j} \]

\noindent in the system $A$. We will now couple the two systems by the von Neumann interaction Hamiltonian  \cite{Aharonov,ACE}:

\begin{eqnarray}
\hat{H}= \hat{H}_{int}= g(t) \hat{A} \hat{P},
\end{eqnarray}

\noindent where $[\hat{X},\hat{P}]=i\hbar$ and the coupling function $g(t)$ satisfies:

\[ \int_0^T g(t)dt = 1 ,\]

\noindent during the coupling time $T$. We shall start with the vector:

\[ \ket{\Psi} = \ket{\psi} \ket{\Phi}, \]

\noindent in the tensor space of the two systems. Applying the time evolution operator we get:

\[ e^{-i\hat{A}\hat{P}/\hbar} \ket{\Psi}.\]

\noindent It is easy to see that on the subspace $\ket{a_j}  V_Q$ the Hamiltonian $\hat{H}$ takes $\hat{X}$ (i.e. $I \cdot \hat{X}$) to $\hat{X}+I a_j$, since by the time $T$ the coupling is already done, we have (Heisenberg equation):

\begin{eqnarray}
\hat{X}(T) - \hat{X}(0) = \int_0^T dt \frac{\partial \hat{X}}{\partial t} = \int_0^T \frac{i}{h}[\hat{H},\hat{X}] dt = a_j,
\end{eqnarray}

\noindent and therefore this Hamiltonian induces a transformation of the operator $\hat{X}$. The corresponding transformation of the coordinates of the wave function is (see \cite{Peres} section 8.4):

\begin{eqnarray}
e^{-i\hat{A}\hat{P}/h}\ket{\psi}\eta(x) = \frac{1}{\sqrt{N}}\sum_j \ket{a_j}\eta(x-a_j).
\end{eqnarray}

We now examine the case of almost discrete pointer by taking:

\[ \ket{k} = \left(\frac{1}{2\pi\epsilon^2}\right)^{1/4} e^{-\frac{(x-k)^2}{4\epsilon^2}}, \]

%\frac{1}{2\pi}^{\frac{1}{4}} e^{-\frac{-(x-k)^2}{2\sigma^2}}\ket{x}\]

\noindent where $\epsilon$ is small enough so that $\ket{k}$ is a sharp Gaussian centered around $x=k$ (this will help us defining an almost discrete function in the variable $k$). Hence,

\[ e^{-i\hat{A}\hat{P}/h}\ket{a_j} \ket{k} = \ket{k-a_j}.\]

\noindent Therefore, for the state vector $\ket{\phi} = \sum \phi(k)\ket{k}$ we have:

\[ e^{-i\hat{A}\hat{P}/h}\ket{a_j} \ket{\phi} = \sum \phi(k) \ket{k-a_j}. \]

\noindent We choose

\[ \hat{A}= \sum_{l=1}^n 2^{l-1} \left({\frac{1-\sigma_z}{2}}\right)^{(l)}. \]

\noindent Let $j=i_n,...,i_1$ be the binary decomposition of $j$:

\[ j= \sum_{l=1}^n i_l 2^{l-1}, \]

\noindent then

\[\hat{A} \ket{i_n \otimes i_{n-1},...\otimes i_1}= j \ket{i_n \otimes i_{n-1},...\otimes i_1}.\]

\noindent Therefore we can write $\ket{a_j} = \ket{i_n \otimes i_{n-1},...\otimes i_1}$,  $a_j =j$, and thus:

\[ e^{-i\hat{A}\hat{P}/h}\ket{a_j} \ket{\phi} = \sum \phi(k) \ket{k-j}. \]

\noindent The above formula is the entanglement we need for the correlation integrals in the main text.

\section {Acknowledgments}

E.C. was supported in part by the Israel Science Foundation Grant No. 1311/14.\\


\section {References}

\begin{thebibliography}{100}

\bibitem{Champeney} D. C. Champeney, Fourier transforms and their physical applications, Academic Press (London, 1973).

\bibitem{Schwartz} M. Schwartz, L. Shaw, Signal processing, discrete spectral analysis, detection and estimation, McGraw-Hill (New-York, 1975).

\bibitem{Sharma} P. D. Sharma, Introduction to modern communication theory, Nem Chand and Bros. (Roorkee, 1971).

\bibitem{Lyons} R. G. Lyons, Understanding digital signal processing, Addison Wesley Pub. Co. (Reading, 2004).

\bibitem{Lathi} B. P. Lathi, Modern digital and analog communication systems, Holt-Saunders international editions (New-York, 1983).

\bibitem{Hancock} J. C. Hancock, An introduction to the principles of communication theory, MacGraw-Hill (New-York, 1961).

\bibitem{Cover} T. M. Cover, J.A.Thomas, Elements of information theory, Wiley (New York, 1991).

\bibitem{Wiener} N. Wiener, Extrapolation, interpolation and smothing of stationary time series, M.I.T. Press (1950).

\bibitem{Garrison} J. C. Garrison, R.Y.Chiao, Quantum Optics, Oxford Univ. Press (2008).

\bibitem{Glauber} R. J. Glauber, Nobel Lecture: One hundred years of light quanta, Rev. Mod. Phys. 78 1267-1278 (2006).

\bibitem{Hanbury} R. H. Brown, The intensity interferometer: its application to astronomy, Taylor and Francis, Ltd.  (New-York, 1974).
    
\bibitem{Gerry} C. C. Gery, P. L. Knight, Introductory quantum optics, Cambridge Univ. Press (2005).

\bibitem{Barnett} S. M. Barnett, Methods in theoretical quantum optics, Oxford Univ. Press (2002).




\bibitem{Neumann} J. von Neumann, Mathematical foundations of quantum mechanics, Princeton Unv. Press (1955).

\bibitem{Aharonov} Y.Aharonov, D.Rohrlich, Quantum paradoxes, Wiley-VCH, (Weinheim, 2005).

\bibitem{ACE} Y. Aharonov, E. Cohen, A. C. Elitzur, Foundations and applications of weak quantum measurements, Phys. Rev. A 89 052105(2014).

\bibitem{Kotler} S. Kotler, N. Akerman, Y. Glickman, A. Keselman, R. Ozeri, Single-ion quantum lock-in amplifier, Nature 473,  61-65 (2011).

\bibitem{Wootters} W. K. Wootters, W. H. Zurek, A single quantum can not be cloned, Nature 299, 802-803 (1982)


\bibitem{Nielsen} M.Nielsen, I.Chuang, Quantum computation and quantum information, Cambrifge Univ. Press (2000).

\bibitem{Amp1} P. B. Dixon, D. J. Starling, A. N. Jordan, J. C. Howell, Ultrasensitive beam deflection measurement via interferometric weak value amplification, Phys. Rev. Lett. 102, 173601 (2009).

\bibitem{Amp2} A. Feizpour, X. Xing, A. M. Steinberg, Amplifying single-photon nonlinearity using weak measurements, Phys. Rev. Lett. 107, 133603 2011.


\bibitem{Peres} A. Peres, Quantum theory: concepts and methods, Kluwer Academic Publishers (Dordrecht, 2002).


%\bibitem{Remark1} One can consider a more general superposition. This will construct a more general cross-correlation integrals.

%\bibitem{Woodward} P. M. Woodward, I. L. Davis, Information theory and inverse probability in telecommunication, Proc. Inst.E.E. 99 (1952).

%\bibitem{Remark2} $p(y/x) = \frac{p(x/y) \cdot p(y)}{p(x)}$






%\bibitem{Bennett1} C.H.Bennett,P.W.Shor Quantum infrmation theory IEEE Trans. Inf. Theory 44 (6) 1998

%\bibitem{Bennett2} C.H.Bennett, D.P.DiVincenzo, Quantum information and computation, Nature 404, 2000

%\bibitem{Holevo} A.S.Holevo, Statistical problems in quantum physics, Proceedings of the second Japan -USSR symposium on probability theory, Springer Verlag, Berlin 1973.

%\bibitem{Schumacher} B.Schumacher, Quantum coding, Phy.Rev.A (51) 1995.

%\bibitem{HSW} B.Schumacher, M.D.Westmoreland, Sending classical information via noisy quantum channel, Phy.Rev.A (56) 1997

%\bibitem{SchuNiels} B.Schumacher, M.A.Nielsen, Quantum data processing and error correction, Phy.Rev.A (54) (4) 1996

%\bibitem{Bruning} E.Bruning, F.Petruccione, (Eds) Theoretical foundation of quantum information processing and communication, Lecture notes in physics vol.787, Springer 2010

%\bibitem{Leuchs} G.Leuchs, T.Beth, (Eds) Quantum information processing, Wiley-VCH,

%\bibitem{Scofield} J.H.Scofield, A frequency domain description of a Lock-in Amplifier, American Journal of Physics 62(2) 1994

%\bibitem{Hays} W.L.Hays, Statistics, Holt, Rinehart and Winston, 1970

%\bibitem{Papageorgiou} A.Papageorgiou, J.F.Traub, Quantum algorithms and complexity for continous problems, arXiv:0712.1211v1 [quant-ph], 2007

%\bibitem{Novac} E.Novac, Quantum complexity of integration, arXive:quantum-ph/0008124 June 2001

%\bibitem{Heinrich} S.Heinrich, E.Novac, Optimal summation and integration by
%deterministic, randomized, and quantum algorithm, arXive:quantum-ph/0105114, May 2001

%\bibitem{Tamir} B.Tamir, S.Masis, Weak measurement and weak information, Foundation of Physics, vol 42, (4), 2012

%\bibitem{Champeney} D.C.Champeney, Fourier transforms and their physical applications, Academic Press, 1973

%\bibitem{Ballard} D.H.Ballard, C.M.Brown, Computer vision, Prentice-Hall, 1982

%\bibitem{Schwartz} M.Schwartz, L.Shaw, Signal processing, Discrete spectral analysis, detection and estimation, MacGraw-Hill, 1975

%\bibitem{Lomont} C.Lomont, Quantum convolution and quantum correlation algorithms are physically impossible, Arxive:quant-ph



%\bibitem{Feller} W.Feller, An introduction to probability theory and its applications, vol II, John Wiley and Sons, 1972

%\bibitem{Peres} A.Peres, Quantum theory: Concepts and Methods, Kluwer Academic Publishers, 2002


%\bibitem{Knuth} D.E.Knuth, The art of computer programming, vol 2/Seminumerical algorithms, Addison-Wesley, 1969



%\bibitem{Remark1} $Var \frac{1}{K}\sum S_j = \frac{1}{K^2} \sum Var(S_j) + \sum_{ij} Cov(\frac{S_i}{K}, \frac{S_j}{K}) = \frac{Var S_j}{K} + \frac{K(K-1)}{K^2} Cov(S_i,S_j)$, where in the last two equalities we assumed the variances and   covariances are the same.




%\bibitem{Papageorgiou} A.Papageorgiou, J.F.Traub, quantum algorithms and complexity for continuous problems, arXiv:0712.1211 [quant-ph]

%\bibitem{Nayak} A.Nayak, F.Wu, The quantum query complexity of approximating the median and related statistics, Proc,STOC 1999, arXiv:quant-ph/9804066

%\bibitem{Piotrowski} E.Piotrowski, J.Sladkowski, An invitation to quantum game theory, arXiv:quant-ph/0211191

%\bibitem{remark1} Use the fact that:

% \[ e^{-2iS_z P_d/nh^2} \frac{1}{\sqrt{2}} (\ket{0}_i+ \ket{1}_i) = \frac{1}{\sqrt{2}} (e^{-i P_d/nh}\ket{0}_i +e^{i P_d/nh}\ket{1}_i) \]

%\bibitem{remark2} Use the fact that $e^{ix} \approx 1+ix$ for $x$ small enough, and this is guaranteed by $n$ big enough.



%\bibitem{Aitchison} B.Aitchison, The Lognormal distribution, Cambridge Univ. Press 1957

%\bibitem{Hays} W.Hays, Statistics, Holt,Rinehart and Winston Inc. 1970

%\bibitem{Papageorgiou} A. Papageorgiou J.F.Traub, Quantum algorithms and complexity for continuous problems, arXiv:0712.1211 [quant-ph]

%\bibitem{Klemens} B.Klemens, Modeling with data: Tools and techniques for scientific computing, Princeton Univ. Press, 2008

\end{thebibliography}
\end{document}